\documentclass[aps,eqsecnum,preprint,floatfix,showpacs,amsmath,amssymb]{revtex4}

\usepackage{graphicx}

\begin{document}

\title{Laser induced reentrant freezing in two-dimensional attractive colloidal systems}

\author{Pinaki Chaudhuri$^{1,2}$, Chinmay Das$^{3}$\\
Chandan Dasgupta$^{1,2}$,
H.R. Krishnamurthy$^{1,2}$
and A. K. Sood$^{1,2}$}

\affiliation{$^1$ Department of Physics, Indian Institute of Science, Bangalore, India,\\
$^2$  Jawaharlal Nehru Centre for Advanced Scientific Research, Bangalore, India,\\
$^3$ Department of Applied Mathematics, University of Leeds, Leeds, UK}

\begin{abstract}
The effects of an externally applied one-dimensional periodic
potential on the freezing/melting behaviour of two-dimensional
systems of colloidal particles with a short-range attractive
interaction are studied using Monte Carlo simulations. In such
systems, incommensuration results when the periodicity of the
external potential does not match the length-scale at which the
minimum of the attractive potential occurs. To study the effects
of this incommensuration, we consider two different models for the
system. Our simulations for both these models show the phenomenon
of reentrant freezing as the strength of the periodic potential is
varied. Our simulations also show that different exotic phases can
form when the strength of the periodic potential is high,
depending on the length-scale at which the minimum of the
attractive pair-potential occurs.
\end{abstract}


\maketitle

\section{Introduction}

In the pioneering experiments of Chowdhury, Ackerson and
Clark\cite{chowdhury} on laser induced freezing, a two-dimensional
monolayer of colloidal particles in the liquid state was subjected
to a laser intensity pattern periodically modulated along one
direction. They found that if the wave-vector of the modulation is
tuned to the wave-vector at which the liquid structure factor
peaks, a triangular lattice with full two-dimensional symmetry
results for laser intensities above a threshold value. In a later
experiment Wei {\it et al} \cite{wei} observed that this
triangular lattice melts if the strength of the laser field is
increased further. Although such a `reentrant melting' phenomenon
was observed in Monte Carlo studies \cite{jay} of colloidal
particles interacting via
Derjaguin-Landau-Verwey-Overbook(DLVO)\cite{dlvo} potential, later
simulations of the same system by Das {\it et al} \cite{das} did
not show the re-entrant phase. However, by extending the
Kosterlitz-Thouless-Halperin-Nelson-Young\cite{kthny} theory of
defect-mediated  melting in two dimensions to the case where an
external periodic potential is present, Frey {\it et
al}\cite{frey} argued that for short-range interactions there will
indeed be a re-entrant transition to a liquid state. Later
experiments\cite{bech} and simulations\cite{strepp2} of charged
colloids seem to confirm the occurrence of re-entrant melting.
Meanwhile, numerical studies\cite{chaudhuri, strepp} of the effect
of the external periodic potential on a system of hard disks have
also shown the occurrence of re-entrant melting.

All the above results correspond to the case when the
commensurability ratio $p=\sqrt{3}a/2d$ (where $a$ is the mean
particle distance and $d$ is the period of the external periodic
potential) has the value of 1, i.e every potential trough is
occupied by particles. In a recent experiment\cite{baug}
corresponding to $p=2$, a new phase - the 'locked smectic state' -
was observed, with the crystalline state being found to melt via
this new intermediate phase. This observation is also in
qualitative agreement with theoretical predictions\cite{frey}.

No study has been reported as yet for the case where there is a
possibility of incommensuration between the periodicity of the
external laser field and some length-scale inherent to the
two-dimensional colloidal system. In the present work we study
just such a case. Specifically, we look at the effect of the
external laser field on a monolayer of colloidal particles, which
interact via a short-range attraction apart from the usual
hard-core repulsive interaction. Such a short-range attraction is
known to arise when hard sphere colloidal particles are mixed with
polymers or smaller colloidal particles, giving rise to an
effective attraction\cite{ao} between the larger particles, called
depletion interaction. It has also been suggested in the context
of colloidal particles confined between two walls\cite{larsen,
grier}, where it is claimed that an attractive minimum in the
effective interaction between a pair of colloidal particles arises
when they are close to the walls\cite{controver}. Effects due to
incommensuration should occur if the length-scale at which the
attractive part of the interaction has its minimum is different
from the periodicity of the externally applied laser field. Novel
phases can then appear in the system depending upon the relative
strengths of the two competing potentials with different length
scales.

Recently, G$\ddot{o}$tze {\it et al} \cite{gotze} studied the
effect of an external potential with one-dimensional periodic
modulation on the liquid-vapour transition line of a
three-dimensional mixture of hard-sphere colloidal particles and
polymers. In their density functional calculations, they obtained
a new {\it stacked fluid} phase, which consists of a periodic
succession of liquid and vapour slabs. The presence of this new
phase is a manifestation of the one-dimensional nature of the
externally applied modulated potential in a three-dimensional
fluid. Their simulations and calculations also showed that the
density profiles exhibit a non-monotonic crossover when the
wavelength of the modulation matches the hard sphere diameter.

In the present work, we analyze the effects of incommensuration on
the phenomenon of freezing/melting of colloidal particles in
two dimensions in the presence of a tunable ``substrate'' or
external potential with one-dimensional periodic modulation. For
this purpose, we consider two different models that are described
in section 2, and use Monte Carlo simulations to study their
behaviour. In our simulations of these two models, we clearly see
that for both low and high values of the external potential the
colloidal particles form crystalline structures, whereas for
intermediate strengths of the external potential the system is in
a modulated liquid phase. We term this phenomenon {\it re-entrant
freezing}. It is also observed that at high values of the external
potential, different kinds of phases can occur depending on the
length-scale at which the attractive part of the inter-particle
potential has its minimum. Our simulation results are described in
detail in section 3. The main conclusions of this study are
summarized in section 4.

\section{The Models}

\subsection{Hard-core particles with short-range attraction}

The generation of an effective attractive interaction between
large hard-sphere colloid particles when they are mixed in a
binary mixture with small hard-sphere colloid particles has been
known for quite some time. In a recent study,
Casta$\tilde{n}$eda-Priego {\it et al} \cite{priego} calculated
this depletion interaction in a strictly two-dimensional binary
mixture of hard disks. Using the expression of the depletion
interaction derived in their work, we have constructed a model for
a two-dimensional system of colloidal particles with short-range
attractive interaction. The pair potential $U(r_{ij})$ between
particles $i$ and $j$ with centre-to-centre distance $r_{ij}$ is
given by:\\
\begin{equation}
U(r_{ij})=
\begin{cases}
\infty  & {\text{if $r_{ij} < \sigma$}}\\
-{\cal U} \Big{[} \cos^{-1}\Big{(}\frac{\eta}{1+\eta}\frac{r_{ij}}{\sigma}\Big{)}-
\Big{(}\frac{\eta}{1+\eta}\frac{r_{ij}}{\sigma}\Big{)} \sqrt{1-\Big{(}\frac{\eta}
{1+\eta}\frac{r_{ij}}{\sigma}\Big{)}^2}
\Big{]} & {\text{if $\sigma \leq r_{ij} \leq \sigma(1+\frac{1}{\eta})$}}\\
0  & {\text{if $r_{ij} > \sigma(1+\frac{1}{\eta})$}}
\end{cases}
\label{ueq}
\end{equation}

Here, $\sigma$ is the particle diameter and ${\cal U}$ and $\eta$ are two
parameters which can be used to tune the depth and width of the
short-range attractive part of $U(r_{ij})$. This interaction
potential has a minimum at $r_{ij}=\sigma$, i.e. when the two
particles touch each other. The strength of the interaction
potential at this minimum is given by
\begin{equation}
U_{min}={\cal U}
\Big{[}
\cos^{-1}
\Big{(}\frac{\eta}{1+\eta}\Big{)}-
\Big{(}\frac{\eta}{1+\eta}\Big{)}
\sqrt{1-\Big{(}\frac{\eta}{1+\eta}\Big{)}^2}
\Big{]}
\end{equation}

In addition, we assume that a particle with co-ordinates $(x,y)$
experiences an external periodic potential of the form
\begin{equation}
V(x,y)={V_e}\cos(\frac{2\pi}{d}y)
\end{equation}
In the above equation, the constant ${d}$ is chosen such that for
a number density of $\phi=N/(L_xL_y)$ ($N$ is the number of
particles and $L_x$, $L_y$ are the lengths of the sides of a
rectangular sample), the modulation is commensurate with a
triangular lattice with nearest neighbour distance
$a_s=1/(\frac{\sqrt{3}}{2}\phi)^{\frac{1}{2}}$, i.e.
${d}=a_s\sqrt{3}/2$. Thus, for the binary mixture we are trying to
model, the wave-vector of the external potential is commensurate
with the smallest reciprocal lattice vector
($q_0=2\pi/(a_s\sqrt{3}/2)$) of the triangular lattice that the
large disks would form at that density in the absence of the
smaller disks. Also, the form of the external potential is such
that its troughs run parallel to the $x$-axis.

For the system under study, the important parameters are
$U_{0}={\beta}U_{min}$, $\beta{V_e}$, $\bar{\rho}=\phi{\sigma^2}$
and $\eta$, where $\beta=1/(k_BT)$, $k_B$ being the Boltzmann
constant and $T$ the temperature. While the attractive part of
$U(r_{ij})$ would like to have the particles touching each other (with inter-particle separation $\sigma$),
the externally applied potential would like to have a density
modulation in the $\hat{\bf y}$ direction with periodicity $d$,
resulting in the incommensuration.

\subsection{Soft-core particles with short-range attraction}

The second model we consider is one where the attractive potential
is considered to be a Gaussian well, with the position of its
minimum being incommensurate with the inter-particle separation.
Thus, the pair potential $U(r_{ij})$ between particles $i$ and $j$
with distance $r_{ij}$ is given by
\begin{eqnarray}
U(r_{ij})&=&\frac{1}{2}
\Big{[}
\frac{(Z^{*}e)^2}{\epsilon}{\Big{(}\frac{{\rm exp}({\kappa}R)}{1+{\kappa}R}\Big{)}^2}
\frac{{\rm exp}({-\kappa}r_{ij})}{r_{ij}} - A{\rm exp}\{-B(r_{ij}-\Lambda)^2\}
\Big{]},
\label{dlvo}
\end{eqnarray}
where $R$ is the radius of the colloidal particles
with surface charge $Z^{*}e$ and $\kappa$ is the inverse of the
Debye screening length. Similarly to the first model, in this case
also a colloidal particle with co-ordinates $(x,y)$ is assumed to
experience an external periodic potential of the form
\begin{equation}
V(x,y)={V_e}\cos(\frac{2\pi}{d}y)
\end{equation}
As in the first model, the modulation of the external potential,
$d$ is chosen such that for colloidal particles with density
$\phi=N/(L_xL_y)$, the modulation is commensurate with a
triangular lattice with nearest neighbour distance
$a_s=1/(\frac{\sqrt{3}}{2}\phi)^{\frac{1}{2}}$, i.e.
${d}=a_s\sqrt{3}/2$. The parameter $\Lambda$ in Eqn.\ref{dlvo},
which determines the position of the minimum of the attractive
part of the potential, is assumed to be incommensurate with the
triangular lattice. The parameter $A$ determines the depth of the
attractive well and $B$ determines its width.

This model was motivated in part by the 
suggestion\cite{larsen, grier} that in a two-dimensional monolayer
of charged colloidal particles in an aqueous solution confined
between two walls, with the `bare' interaction between
pairs of particles being given by the
Derjaguin-Landau-Verwey-Overbook(DLVO) potential\cite{dlvo}, an
attractive well with a depth of the order of $k_BT$ develops in
the {\it effective potential} between a pair when the particles
are near a charged wall. The depth and the position of the minimum
are supposed to be strongly dependent on the distance from the
wall, the effective charge on the colloid particles, the
counter-ion density and the surface charge on the wall. However,
this claim has been contested\cite{controver}. In any case,
leaving aside the question of a physical realization of the above
model, its use tests the robustness of our results with respect to
the details of the model interaction.

\section{Simulational details and Results}

\subsection{Model 1}

We have carried out Monte Carlo simulations of a system of $N$
particles of diameter $\sigma$, interacting via the potential
$U(r_{ij})$ defined in Eqn.\ref{ueq}. The particles are contained
in a rectangular box of dimension $L_x \times L_y$, where
($L_y/L_x=\sqrt{3}/2$), with periodic boundary condition being
used for doing the Monte Carlo simulations. Most results reported here are for
simulations done for $N=1600$ particles. In order to check finite
size effects, we will also discuss results for $N=1024$ and $900$.

\begin{figure}[htbp]
\includegraphics[height=8cm,width=8cm]{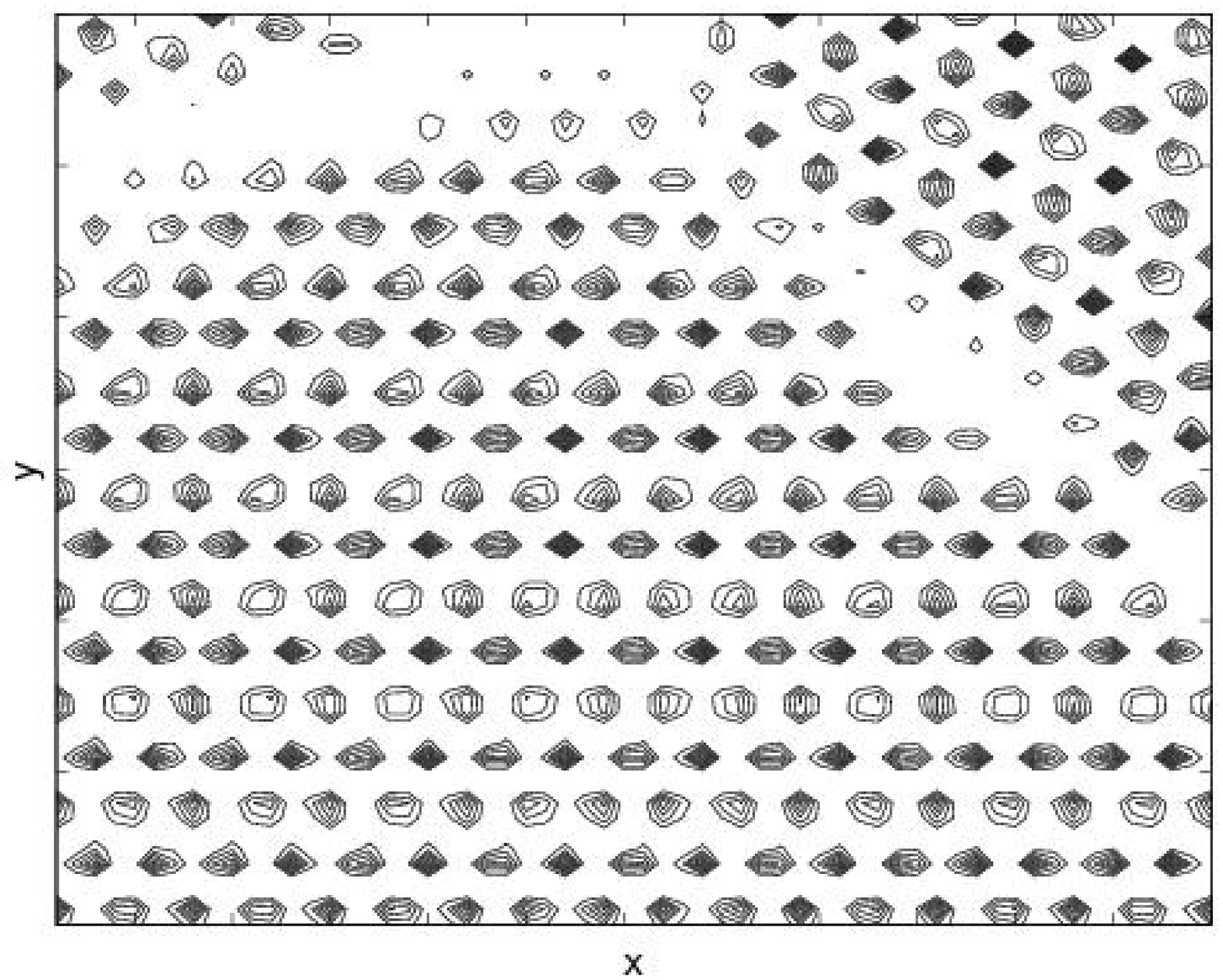} \\
\caption{Model 1. Average particle density $\rho(x,y)$
at ${\bar{\rho}}=0.86$, for the case when the external field is absent,
showing co-existence of crystalline clusters with a gaseous phase.}
\label{fig1a}
\end{figure}

\begin{figure}[htbp]
\includegraphics[scale=0.6,clip=]{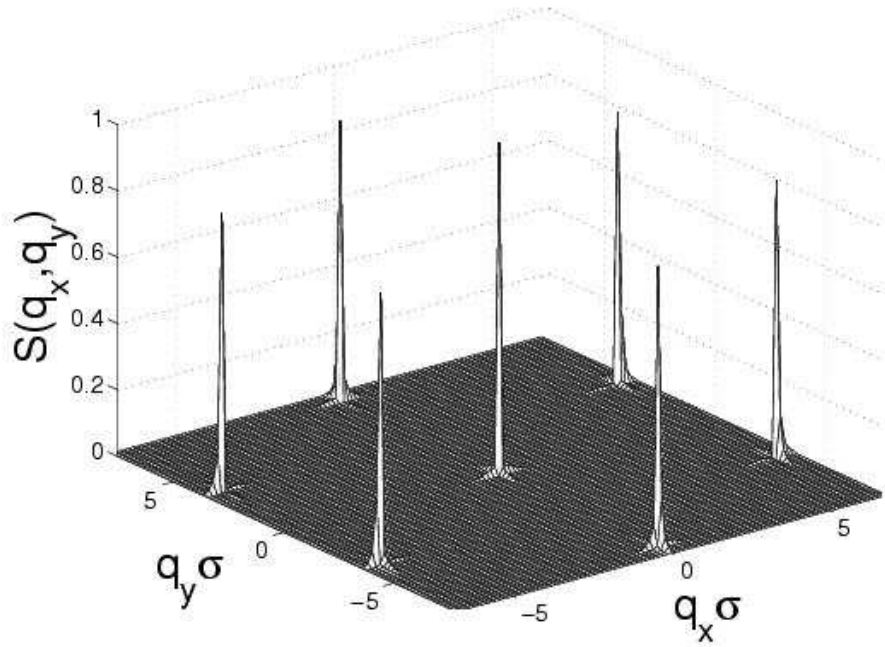} \\
\caption{Model 1. Struture factor $S({\bf q})$ of a colloidal cluster
at ${\bar{\rho}}=0.86$, for the case when the external field is absent,
showing the existence of triangular lattice structure within the cluster.}
\label{fig1b}
\end{figure}

For a system of particles interacting via the potential specified
in Eqn.\ref{ueq}, the phase diagram is not known even when the
external laser field is absent . However, Brownian dynamics
simulations for a similar system\cite{cerda} suggest that at $U_0
\approx 3.1$, there is a transition from a single, dispersed phase
to a phase where the colloidal particles start forming clusters.
In our present work, we consider the case $U_0 = 5.4$ and
$\eta=30$. The first parameter defines the depth of the attractive
potential and the second one fixes its width ($\sigma/30$). In the
range of densities of our interest, $0.85 < \bar{\rho} < 0.95$,
Bolhuis {\it et al} \cite{bolhuis} have observed the co-existence
of a high-density solid with a dilute gas for a two-dimensional
system of particles interacting via a short-range attractive
square potential. Co-existing gas and solid phases are also found
in three dimensions \cite{anderson} when the inter-particle
attraction has a short range. Our simulations also show the
formation of clusters for the choice of $U_0$ and $\eta$ mentioned
above. In Fig.\ref{fig1a}, we have plotted the average density,
representative of our simulated system, at $\bar{\rho}=0.86$. The
plot clearly shows the co-existence of crystalline clusters with a
gaseous phase, with particles arranged in a triangular lattice
inside the clusters, as observed in the Brownian dynamics
simulations\cite{cerda}. The structure factor, $S({\bf q})$, for a
cluster, plotted in Fig.\ref{fig1b}, shows peaks corresponding to
a triangular lattice structure with spacing equal to $\sigma$, the
hard sphere diameter.

Our objective is to study how such a system of colloidal particles
behaves in the presence of an external potential with
one-dimensional periodic modulation $d$ which is not commensurate
with the distance between the lattice planes of the triangular
structure that the particles form in the absence of the field. For
this purpose we calculate the average density $\rho(x,y)$ and the
structure factor $S({\bf q})$ for different strengths of the
potential at a fixed particle density.

\begin{figure}[htbp]
\includegraphics[height=16cm,width=16cm]{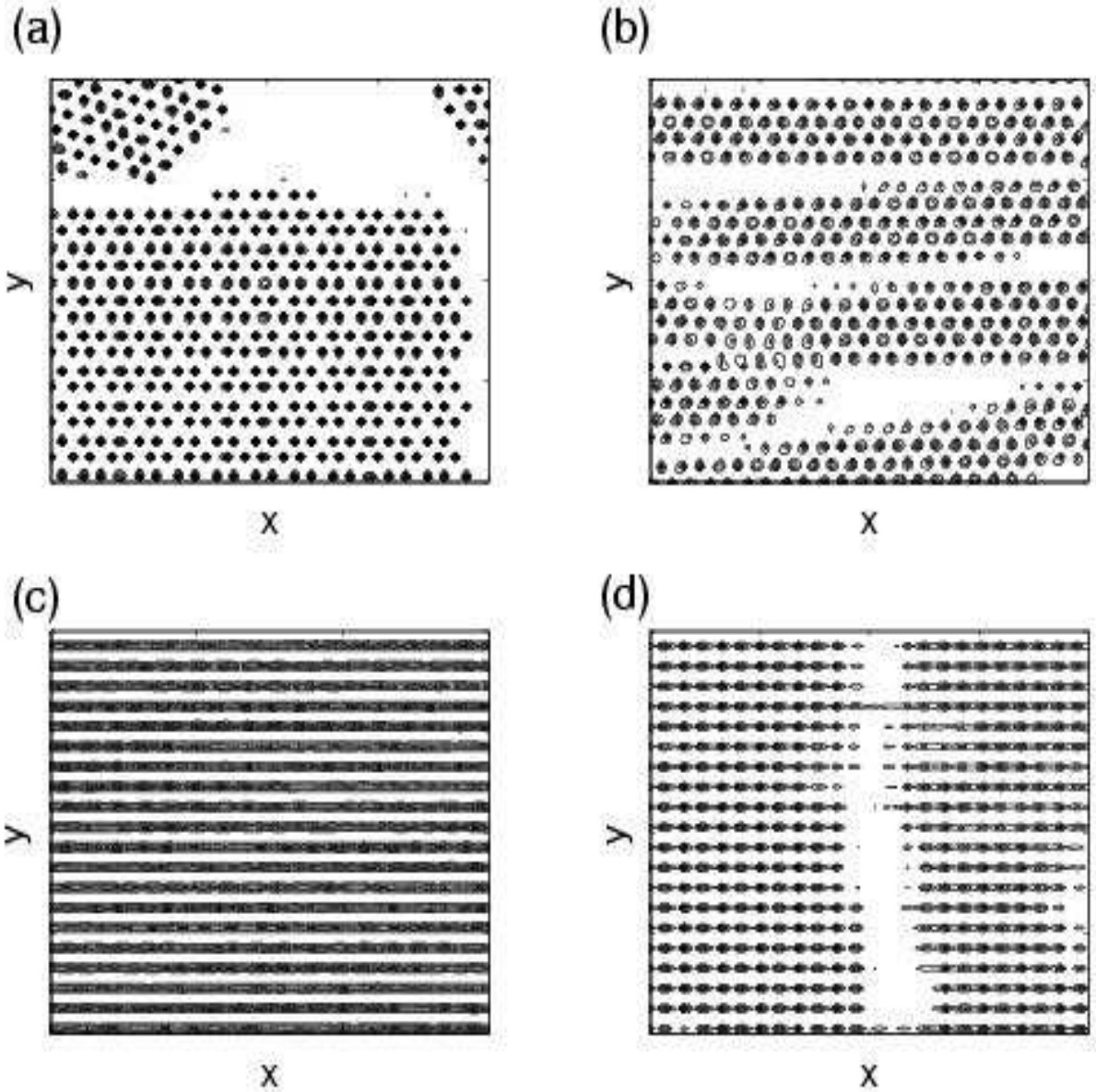} \\
\caption{Model 1. At ${\bar{\rho}}=0.86$, the average particle density $\rho(x,y)$
shows (a) 2d triangular lattice when ${\beta}V_e=0.5$, (b) 'frustrated' liquid when ${\beta}V_e=2.0$,
(c) modulated liquid when ${\beta}V_e=5.0$, (d) 2d square lattice ${\beta}V_e=100.0$. Thus, at this
density, the system undergoes re-entrant freezing as a function of the strength of the 
externally applied field.}
\label{fig2}
\end{figure}

\begin{figure}[htbp]
\includegraphics[scale=0.8]{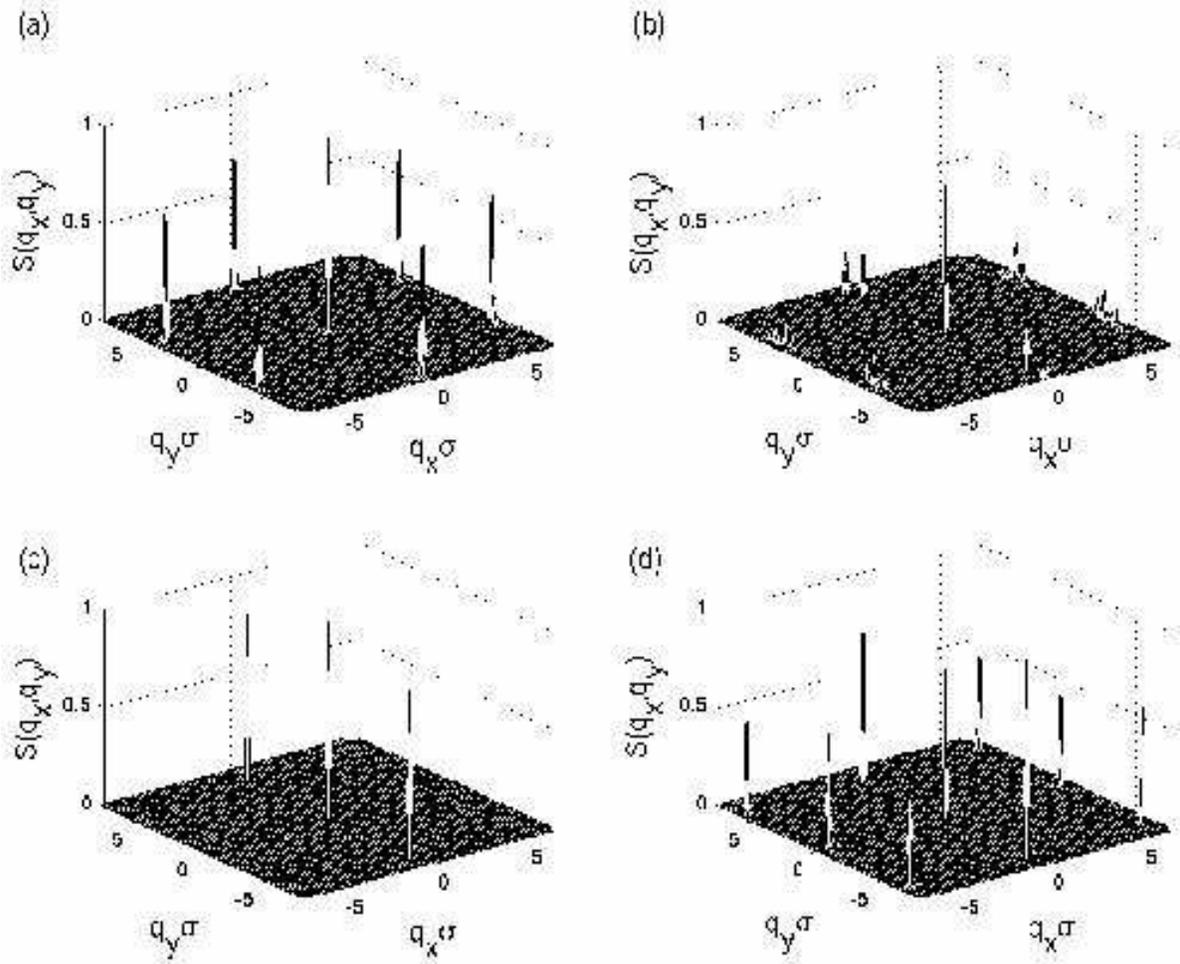} \\
\caption{Model 1. For ${\bar{\rho}}=0.86$, Struture factor $S({\bf q})$
shows (a) 2d triangular lattice when ${\beta}V_e=0.5$, (b) 'frustrated' liquid when ${\beta}V_e=2.0$, 
(c) modulated liquid when ${\beta}V_e=5.0$, (d) 2d square lattice when ${\beta}V_e=100.0$,
confirming the re-entrant freezing experienced by the system as ${\beta}V_e$ is increased. }
\label{fig3}
\end{figure}

For $\bar{\rho}=0.86$, when the strength of the potential is low,
our simulations show that the system continues to form crystalline
clusters with inter-particle separation $\sigma$. In
Fig.~\ref{fig2}(a) and Fig~\ref{fig3}(a) , we have plotted
respectively the average density and the corresponding $S({\bf
q})$ for the colloidal system when the strength of the
laser-induced potential is ${\beta}V_e=0.50$. As can be seen from
the plots, the nature of $S({\bf q})$ is similar to the case where
the potential is absent.

When the strength of the potential is increased to
${\beta}V_e=2.0$, the clusters break up and the particles try to
align themselves along the potential troughs. The average density
$\rho(x,y)$ for this value of ${\beta}V_e$, plotted in
Fig.\ref{fig2}(b), shows that although there is a local triangular
structure, the particles are also getting arranged in the
$\hat{\bf y}$ direction, due to the influence of the external
potential. The $S({\bf q})$ for this situation, plotted in
Fig.\ref{fig3}(b), shows that the height of the peaks
corresponding to  the triangular lattice has decreased
considerably compared to that in Fig.\ref{fig3}(a) and new peaks
have begun to appear on the $\hat{\bf{q}_y}$ axis with $q_y$
values corresponding to the wave-vector of the applied field. This
clearly is an effect of the incommensuration. The system is in a
`frustrated' state -- the particles are making an attempt to lie
at the troughs of the periodic potential, but the energy they gain
by doing so is not sufficient to break the `bonds' of the
triangular lattice with spacing $\sigma$.

In Fig\ref{fig2}(c), we have plotted $\rho(x,y)$ for the system
when the strength of the laser field has been increased to ${\beta}V_e=5.0$. The
crystalline clusters have now melted - the particles have become
confined along the potential troughs. The $S({\bf q})$ for the
system, plotted in Fig\ref{fig3}(c), is characteristic of a
modulated liquid, with the peaks of the structure factor located
only at multiples of the characteristic wave-vector $q_y =
2{\pi}/d$ of the external periodic potential.

When the strength of the periodic potential is increased to higher
values, the motion of the particles in the direction transverse to
the potential troughs decreases considerably. The average density
for the particles at a potential strength of ${\beta}V_e=100.0$,
plotted in Fig.\ref{fig2}(d), shows crystalline order
corresponding to a square lattice, which is also reflected in the
corresponding $S({\bf q})$,
shown in Fig.\ref{fig3}(d). The peaks of the structure factor on
the $\hat{\bf{q}_y}$ axis occur at wave vectors that correspond to
multiples of the wave-vector of the external potential, whereas
the peaks on the $\hat{\bf{q}_x}$ axis correspond to an
inter-particle separation of $\sigma$ inside each potential
trough.

\begin{figure}[htbp]
\includegraphics[height=15cm,width=15cm]{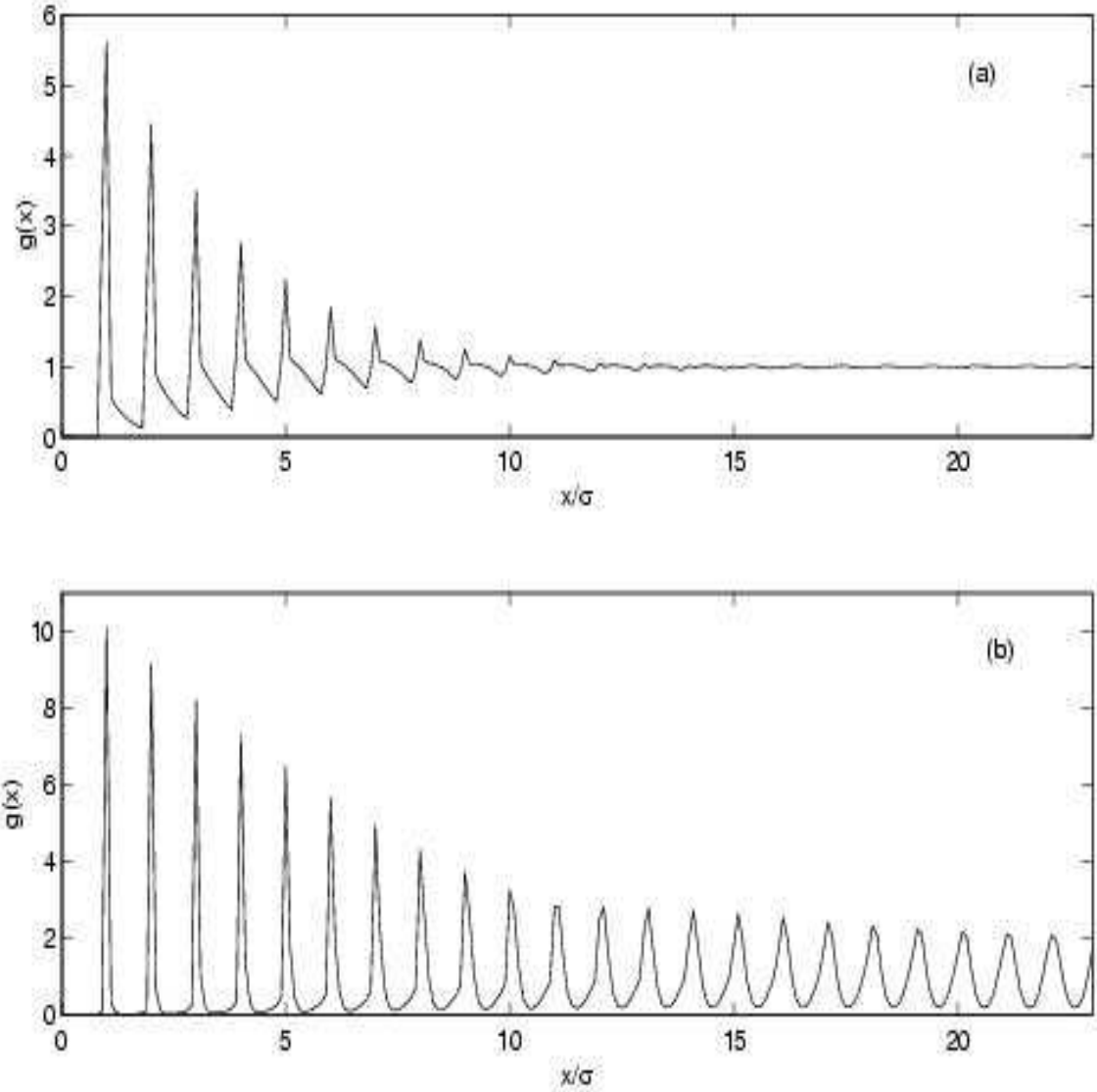} \\
\caption{Model 1. Pair correlation functions $g(x)$, at ${\bar{\rho}}=0.86$, along the potential troughs for (a) modulated
liquid (when ${\beta}V_e=5$) and (b) square lattice (when ${\beta}V_e=100$).}
\label{fig4}
\end{figure}

We have also computed the spatial correlation function $g(x)$,
which is the pair correlation function along the potential
troughs, for the two phases observed for the potential strengths
${\beta}V_e=5.0$ and ${\beta}V_e=100.0$. The maxima of the
correlation function for ${\beta}V_e=5.0$, plotted in
Fig\ref{fig4}(a), decay quite fast, and at long distances $g(x)
{\rightarrow} 1$. This is characteristic of liquid-like behaviour
for the particles confined in the potential troughs. However, when
${\beta}V_e=100.0$, from the $g(x)$ data plotted in
Fig\ref{fig4}(b), we can conclude that there is a periodic
modulation of the density of the colloidal particle in the
potential troughs, the period of this modulation being $\sigma$ as
expected.

To check whether the formation of the crystalline solid at the
higher values of ${\beta}V_e$ is a finite size effect, we have
calculated the corresponding order parameter $\rho_{\bf q}$ for
different system sizes, $N=1600, 1024, 900$. For the solid phase,
the order parameters are the Fourier components of the density
$\rho({\bf r})$ calculated at the reciprocal lattice points
$\{{\bf q}\}$. From the peaks of the structure factor plotted in
Fig.\ref{fig3}(d), we can get the four smallest reciprocal lattice
vectors for the square lattice formed by the particles for
${\beta}V_e=100.0$. Of these four vectors, the two lying on the
$\hat{\bf{q}_y}$-axis correspond to the ordering due to the
external field and the $\rho_{\bf q}$ for these vectors are also
non-zero for the modulated liquid. So the $\rho_{\bf q}$ for the
other two wave-vectors, denoted by $\rho_{\bf q}^{\perp}$, are the
relevant order-parameters for checking the effects of the
finiteness of the system. The average order parameter $<\rho_{\bf
q}^{\perp}>$, has the values $0.731, 0.663, 0.661$ for
$N=900,1024,1600$ respectively, implying weak dependence on the
system size. Therefore, we can conclude that the square lattice
formed at ${\beta}V_e=100.0$ represents a crystalline phase.

The appearance of this crystalline phase at a large value of
${\beta}V_e$ can be understood from the fact that, for
$\bar{\rho}=0.86$, the inter-trough spacing, which is governed by
the periodicity of the external field $d$, is approximately equal
to $\sigma$, the distance at which $U(r_{ij})$ has its minimum. In
fact, for $0.811 \leq \bar{\rho} \leq 0.866$, the spacing $d$ is
within the range of the attractive part of the potential, $1.033
\leq r/\sigma \leq 1.0$, for the value of $\eta$ used in our
simulations. Thus, at $\bar{\rho}=0.86$, if the transverse motion
within a trough is suppressed, correlations develop across the
troughs resulting in ordering of the particles in the $\hat{{\bf
y}}$ direction, in addition to the ordering with spacing $\sigma$
along the troughs. Such a crystal structure corresponds to the
lowest energy configuration consistent with the density and the
inter-trough spacing.

Thus the system of colloidal particles, which had preferred to
form crystalline clusters at low field strengths, regains a
different crystalline phase at high field strengths, after passing
through an intermediate modulated liquid phase. This phenomenon
may be called {\it re-entrant freezing}. It is interesting to note
that the two crystalline phases have different symmetries, which
is a consequence of the incommensuration effects. Another aspect to
note is that the solid is in co-existence with a dilute gas and
the `voids' corresponding to the gaseous phase can occur at any
point in space. This is a probable reason for the decrease in the
height of the maxima of $g(x)$ at large distances. The
inter-trough correlations, observed at ${\beta}V_e=100.0$, are not
developed at lower field strengths because the motion of the
particles in the transverse direction is not suppressed
sufficiently and the particles can vibrate within the width
provided by the trough, thereby causing a destruction of the
crystalline order.

However, we should note that the kinetics of systems with
hard-core interactions may depend strongly on the initial
state\cite{anderson}. In our simulations, we observe such
behaviour. For some initial configurations, the system gets
kinetically jammed and due to this jamming, even at high field
strengths, there is a co-existence of the square and triangular
phases. In simulations, one can escape from jamming if a
combination of local and non-local Monte Carlo moves are used.

\begin{figure}[htbp]
\begin{tabular}{l}
(a) \\
\centerline{\includegraphics[height=8cm,width=8cm]{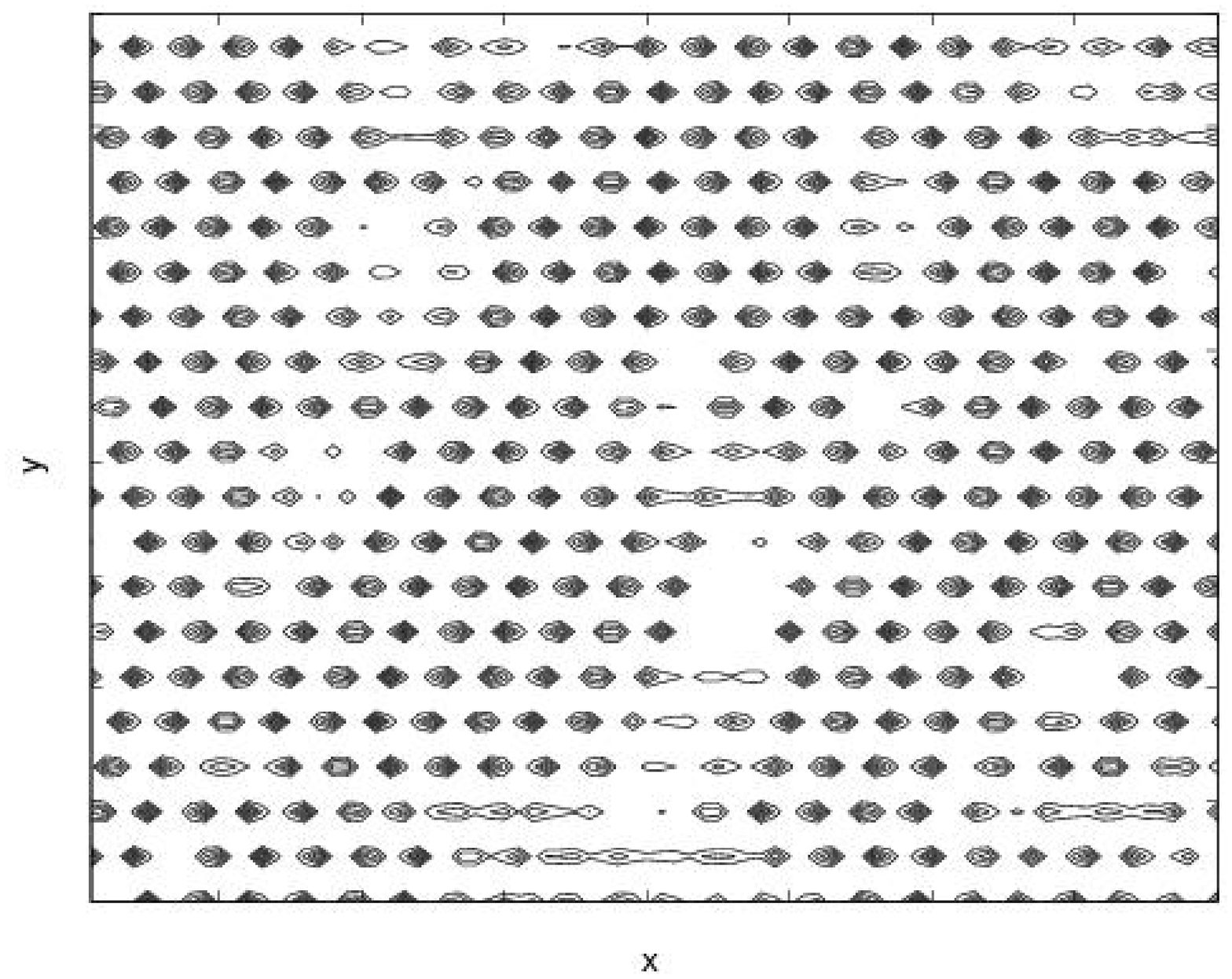}} \\
(b) \\
\includegraphics[angle=-90,scale=0.85]{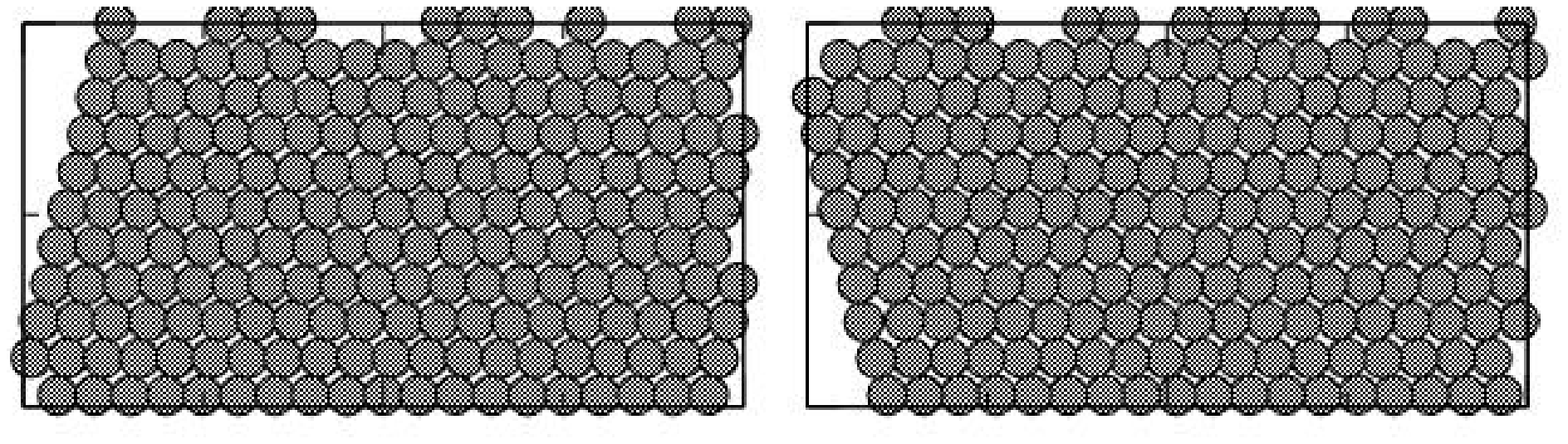} \\
\end{tabular}
\vspace*{-5in}
\caption{Model 1. (a).The average density for $\bar{\rho}=0.92$ at ${\beta}V_e=100$, showing that the 
particles form a mixed crystal.\\ (b). The two panels
are snapshots of the crystalline structures having same energy, one of which will be formed by the particles
as ${\beta}V_e \rightarrow \infty$, at $\bar{\rho}=0.92$.}
\label{fig5}
\end{figure}

For $\bar{\rho} > 0.866$, the wavelength of the external potential
becomes smaller than the particle diameter and therefore, for
these densities, the square lattice become unfeasible at high
potential strengths. However, even for these densities, the
particles in neighbouring troughs would like to be in contact with
each other and that can only happen if they form a rhombic
structure. In Fig.\ref{fig5}, we have plotted the average density
for $\bar{\rho}=0.92$ at ${\beta}V_e=100.0$. The particles have
formed a `mixed crystal' , i.e they have formed domains which have
local crystalline structure similar to the two structures shown in
the figures below the $\rho(x,y)$ plot. As ${\beta}V_e \rightarrow
\infty$, the motion of the particles becomes one-dimensional and
the system tries to attain one of these two crystalline structures
in order to minimize its free energy.

\subsection{Model 2}

For the other choice of the pair potential given in
Eqn.\ref{dlvo}, we have considered $400$ particles in a
rectangular box commensurate with a triangular lattice structure
and with periodic boundary conditions. The screening length is
fixed at ${\kappa}a_s=15.0$ where $a_s$, the lattice spacing in a
triangular arrangement, has been used as the unit of length in all
the expressions. Without any external field, the system (colloidal
particles of diameter $1.07{\mu}m$, surface charge $Z^{*}=7800e$
and density $n_p=1.81\times10^7/cm^2$, suspended in water having
dielectric constant $\epsilon=78$ at a room temperature of $298K$)
is in the liquid phase. The parameters for the Gaussian well are
chosen as $A=\frac{V_{e}}{2}$ and $B=50$. In the first part of our
simulations for this system, the position of the attractive
minimum is set at $\Lambda=1.3$.

\begin{figure}[htbp]
\includegraphics[angle=90,width=12cm, height=12cm,clip=]{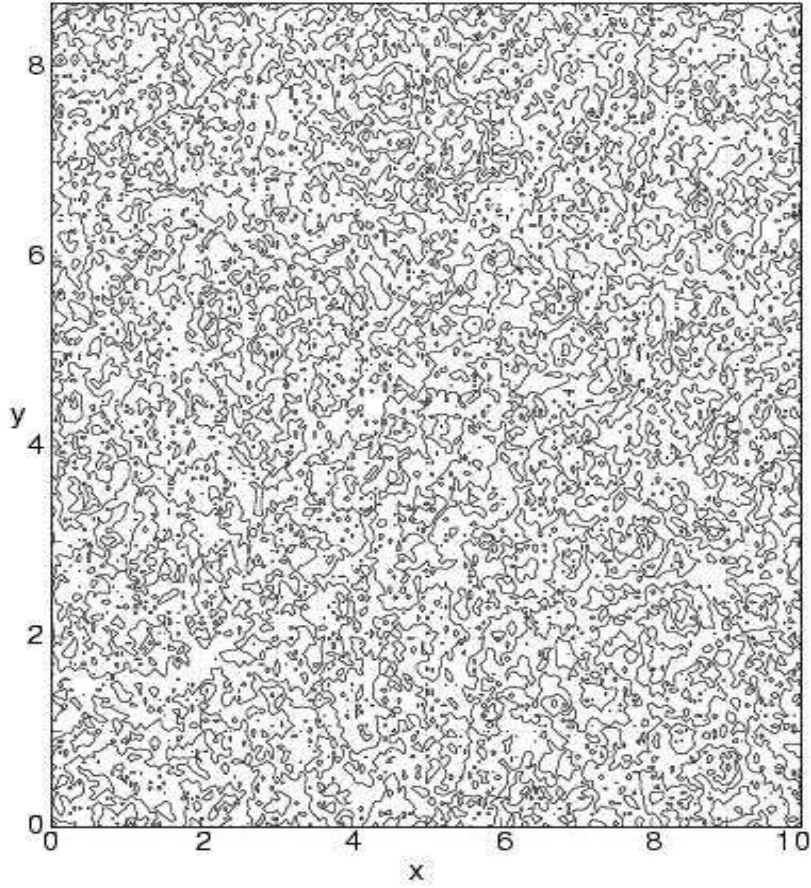}
\caption{ Model 2. Average density $\rho(x,y)$ in the absence of the external field and 
attractive potential for ${\kappa}a_s=15$.}
\label{ch1}
\end{figure}

To observe the effect of the laser field in this system of
particles, the average density is calculated for different values
of ${\beta}V_e$. In Fig.\ref{ch1}, we have shown a contour plot of
the average density, representing a section of the simulation box,
in the absence of the external potential. The average density does
not show any order, signifying a liquid state.

\begin{center}
\begin{figure}[htbp]
\caption{Model 2. The left hand panel (a, c and e) shows the average density for simulation with just the DLVO potential. The
right hand panel (b, d and f) shows the same for simulation with attractive potential.}
\includegraphics[angle=0,scale=1.0,clip=]{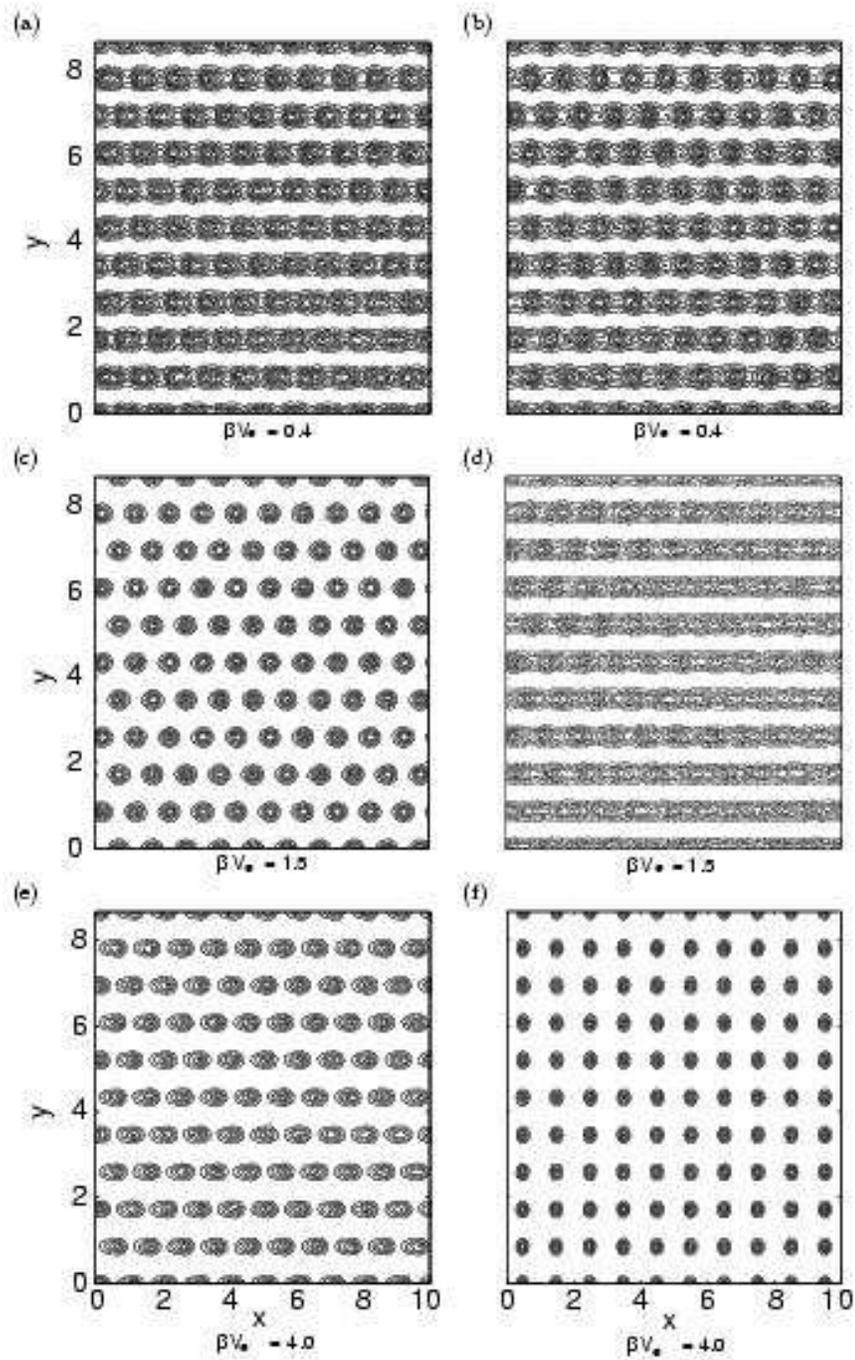} 
\label{ch2}
\end{figure}
\end{center}

Fig.\ref{ch2} contrasts the effect the external potential has on
the average density depending on whether the inter-particle
potential has the attractive part or not. At ${\beta}V_e=0.4$
(subplots (a) and (b)), the liquid-like structure of Fig.\ref{ch1}
is replaced by one corresponding to a triangular lattice structure
in both cases. The order parameter $\rho_{\bf q}^{\perp}$, corresponding to the
triangular lattice structure, has the value $0.6$ for the pure DLVO potential
and $0.53$ for the case when the attractive part is present.
As the field is increased to ${\beta}V_e=1.5$, the
crystalline order for the pure DLVO potential becomes much
sharper, see Fig.\ref{ch2}(c), and the value of $\rho_{\bf q}^{\perp}$ (for 
the triangular lattice) becomes
higher. But with the attractive part
present (see Fig.\ref{ch2}(d)), the average density looks like a
set of liquid-like strings, indicating a modulated liquid phase, with 
$\rho_{\bf q}^{\perp}$(for the triangular lattice) becoming much smaller $\approx 0.15$.
At still higher fields, ${\beta}V_e=4.0$, nothing changes
qualitatively for the pure DLVO case (see Fig.\ref{ch2}(e)). However, the average density
plot for the case with the attractive potential (Fig.\ref{ch2}(f)) indicates the
formation of a rectangular lattice.

Therefore, for this choice of inter-particle potential, we again
find that the system of colloidal particles, which had formed a
crystalline structure for low strength of the external potential,
melts as the value of ${\beta}V_e$ is increased. But at stronger
external potential, the particles again form a crystalline
structure which has rectangular symmetry, similar to what we had
observed for the case of Model-$1$. Hence, here also, we observe a
re-entrant crystallization.

\begin{figure}[htbp]
\caption{Model 2. Average density for ${\kappa}a_s=15$, ${\beta}V_e=4$ and with different values of $\Lambda$.}
\label{fig:lim:lambda}
\includegraphics[angle=0,scale=1.0,clip=]{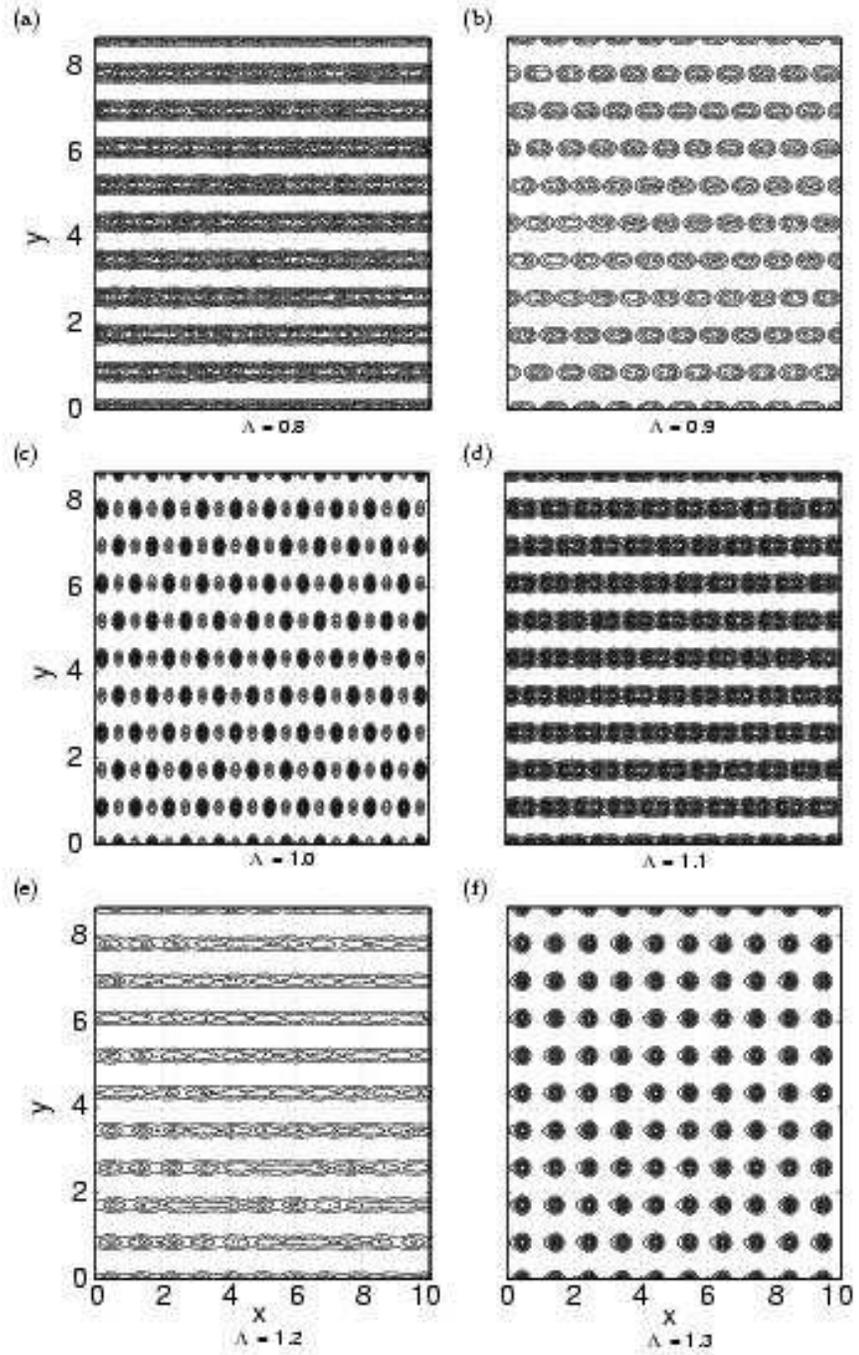}
\vspace{-4cm}
\end{figure}

\begin{center}
\begin{figure}[htbp]
\begin{tabular}{l l}
(a) & (b) \\
\includegraphics[scale=0.4,clip=]{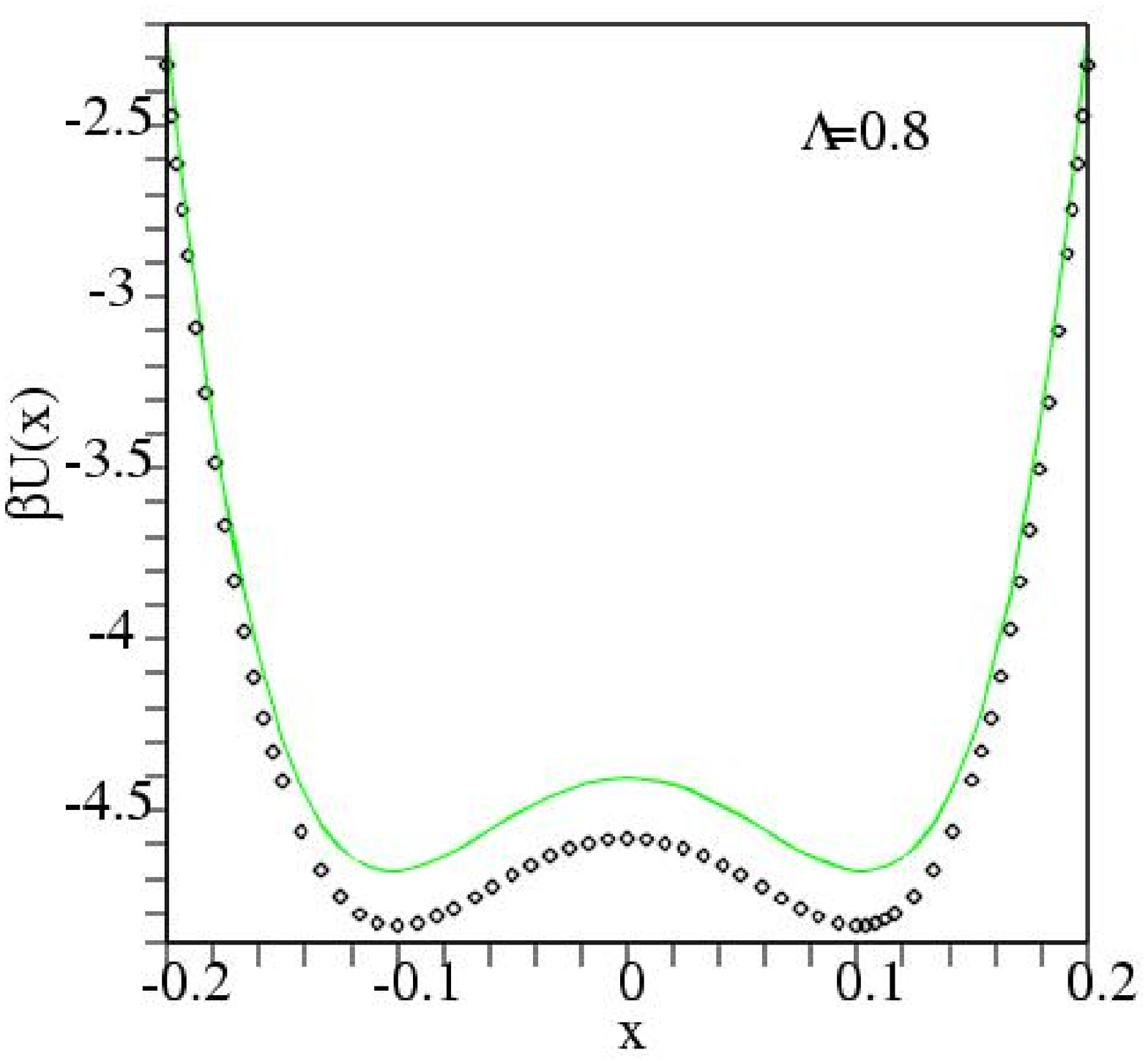} &
\includegraphics[scale=0.4,clip=]{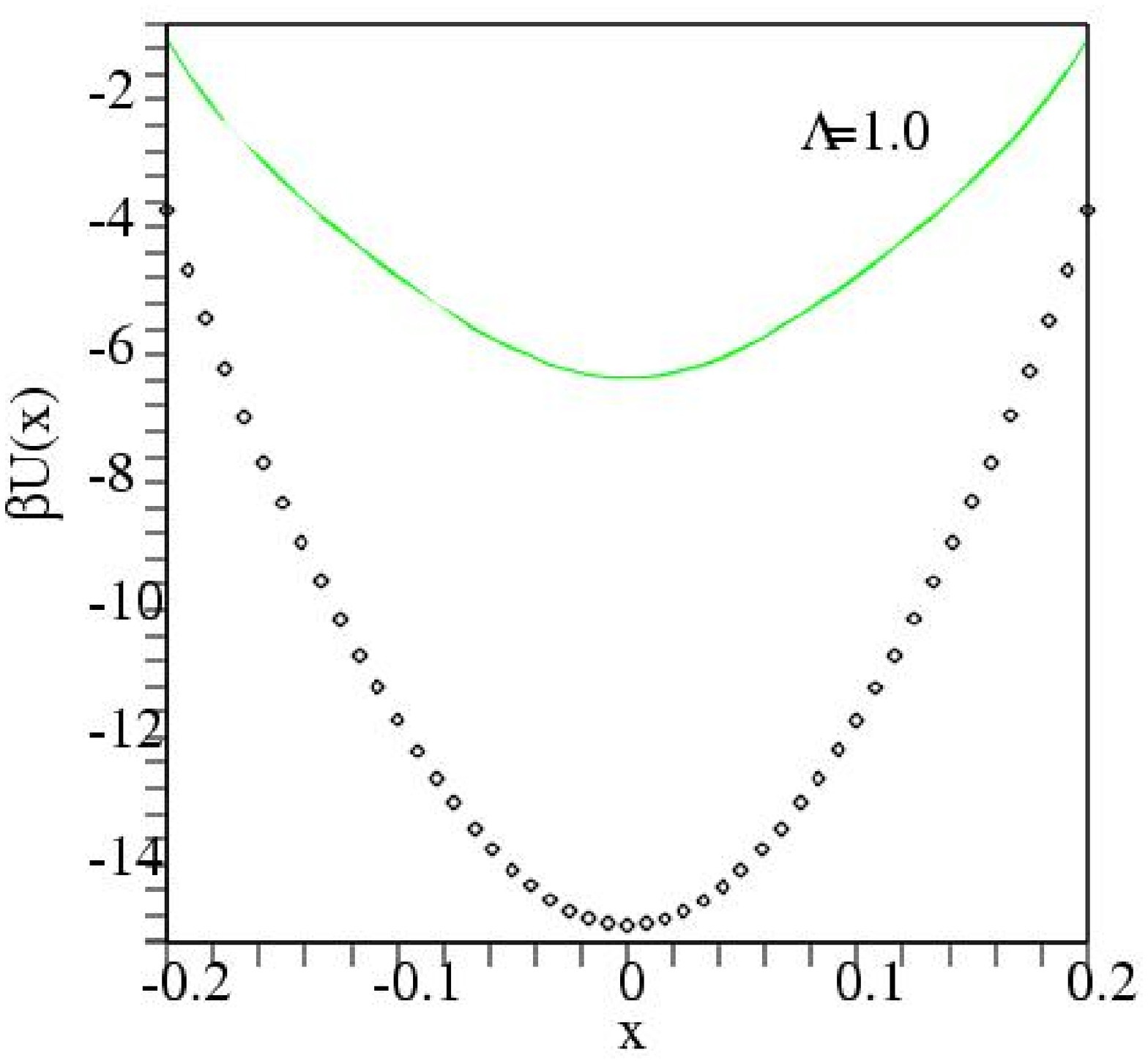} \\
(c) & (d) \\
\includegraphics[scale=0.4,clip=]{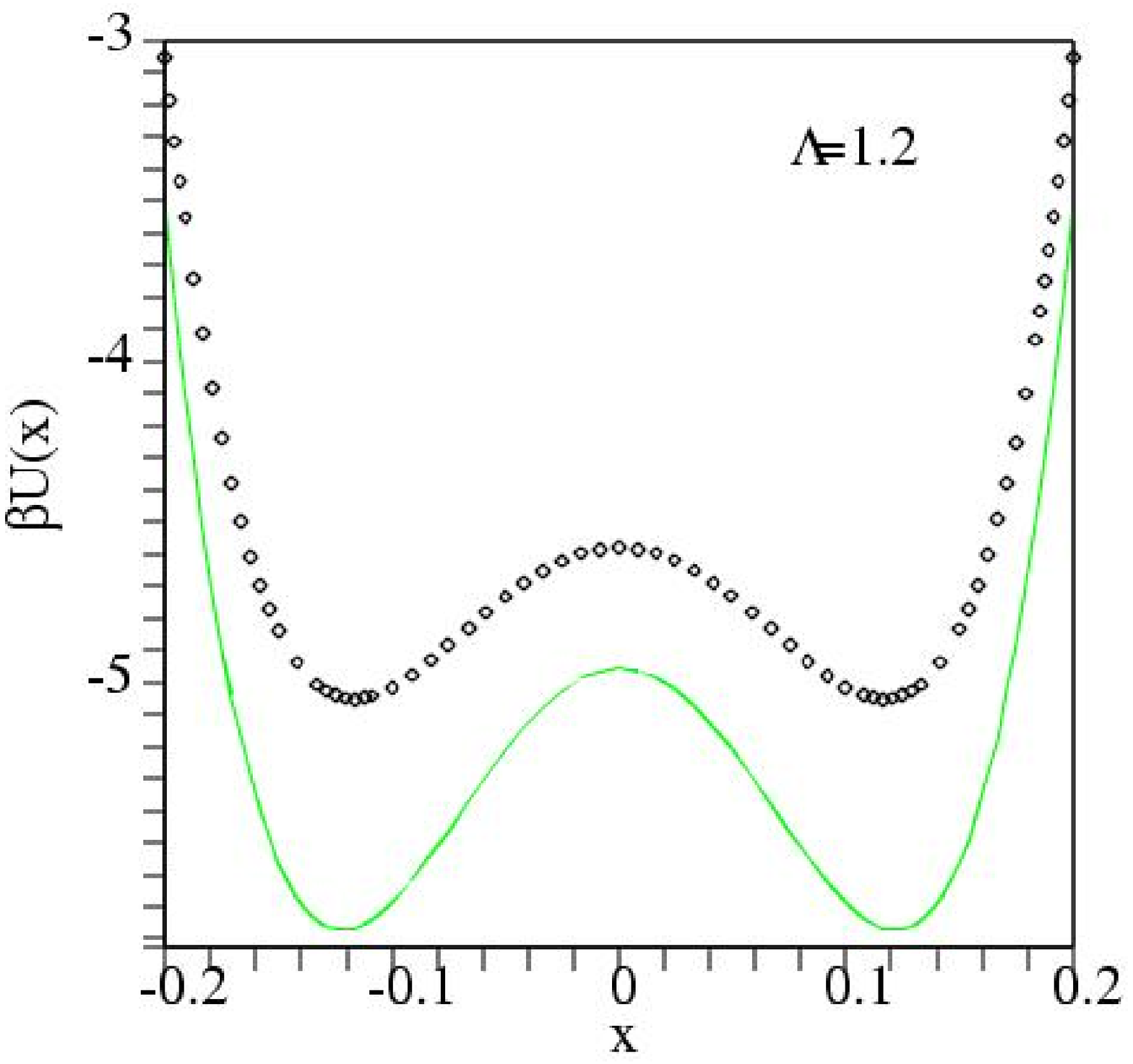} &
\includegraphics[scale=0.4,clip=]{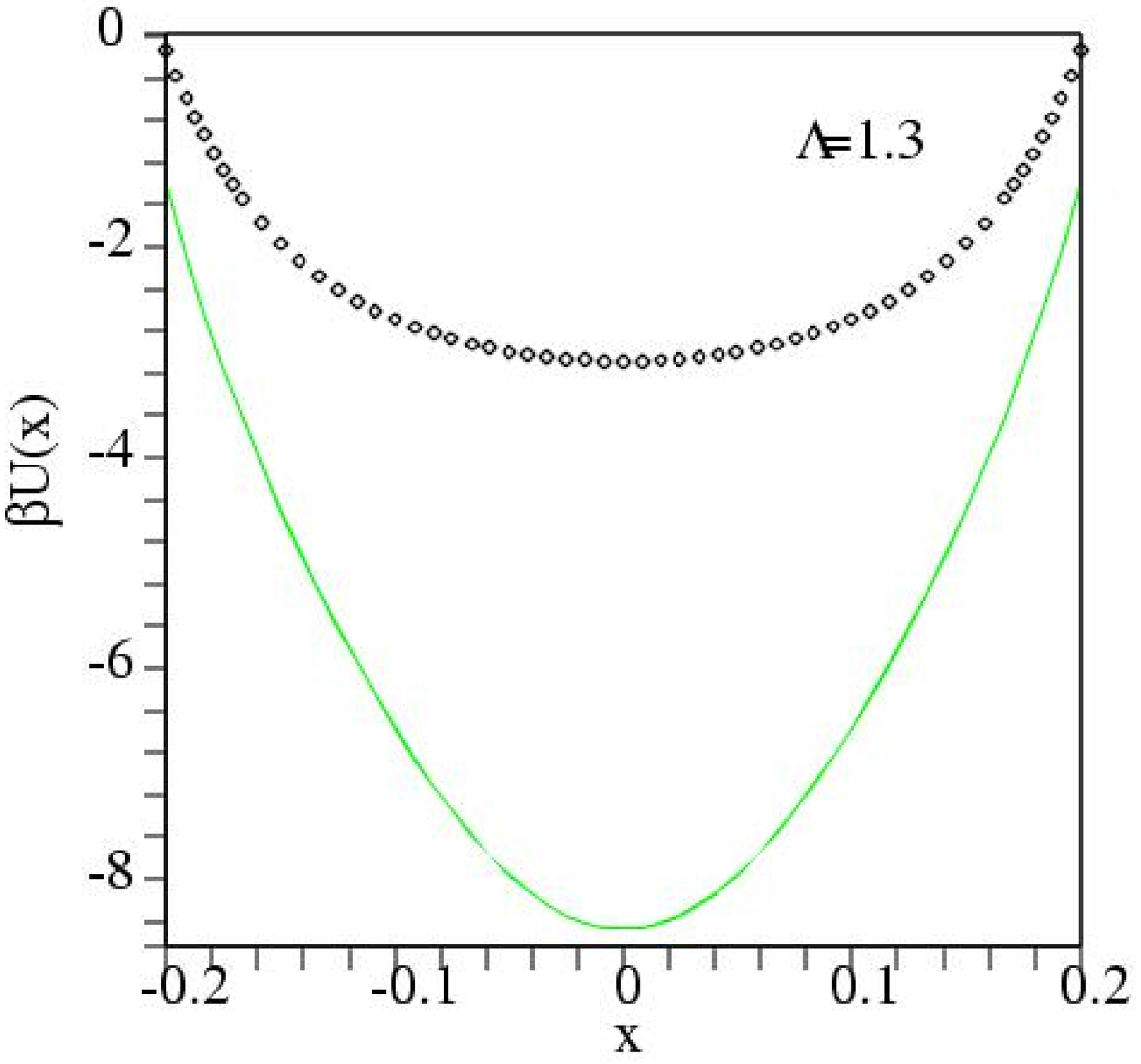} \\
\end{tabular}
\caption{Model 2. Potential experienced by a given particle with neighbours in perfect
hexagonal (points) or rectangular (lines) arrangement for different values
of $\Lambda$.}
\label{fig:lim:pot}
\end{figure}
\end{center}

At high external potential strengths, different choices of the
parameter $\Lambda$ need not always result in a rectangular
lattice. In Fig.~\ref{fig:lim:lambda} we have plotted the average
density for various other values of $\Lambda$ and ${\beta}V_e =
4.0$. It shows that a rich variety of phases can arise. To
understand the structures, we consider the potential energy of a
given particle in either triangular or rectangular lattice
arrangement. Since both the DLVO and the Gaussian attractive part
of the potential fall off rapidly with increasing particle
separation, the contribution from the first shell of neighbours is
the most significant. Because of the large ${\beta}V_e$, the
potential rises sharply in the $y$ direction from the potential
troughs. Thus we look for the potential experienced by a particle
along the $x$ direction with the $y$ co-ordinate fixed to be at a
minimum of the external potential.  Considering the six nearest
neighbours for the triangular lattice case and the eight nearest
and next-nearest neighbours for the rectangular lattice, we have
calculated the effective potential well when all the neighbours
are in perfect lattice positions (Fig.~\ref{fig:lim:pot}). For
$\Lambda=0.8$, the average density (Fig.~\ref{fig:lim:lambda}(a))
looks like a modulated liquid with some superposed modulation
perpendicular to the field direction also.
Fig.~\ref{fig:lim:pot}(a) shows the behaviour of the potential
well for the same parameter values -- both triangular and
rectangular lattices are energetically {\em unstable}. The energy
minimum is for a hexagonal arrangement of the neighbours similar
to that in triangular lattice, but the
central particle has two equivalent energy minima displaced from
the center. Thus the system remains frustrated with some residual
hexagonal order.

As $\Lambda$ is tuned to a value of $1.0$, a deep minimum in the
single particle potential corresponding to a triangular lattice
develops (Fig.~\ref{fig:lim:pot}(b)). However, since we increased
$\Lambda$ through a frustrated potential structure, the system
takes a long time to relax to an unique triangular lattice
spanning the whole system. Fig.~\ref{fig:lim:lambda}(c) shows the
average density plot for $\Lambda = 1.0$. To accommodate the
history of the frustrated structure at lower $\Lambda$, the shown
part of the system goes through a coherent shift along the $x$
direction. At long times, the average density shows sharp contours
corresponding to a triangular lattice.

With increasing $\Lambda$, the single particle potential with the
neighbours in a regular lattice structure again shows degenerate
minima, but this time the rectangular lattice wins over the
triangular lattice energetically. For $\Lambda=1.2$ the single
particle potential is shown in Fig.~\ref{fig:lim:pot}(c) and the
corresponding average density contours in
fig.~\ref{fig:lim:lambda}(e). The structure is that of modulated
liquid, with superimposed rectangular modulation.

When $\Lambda = 1.3$, the rectangular structure is energetically
stable and more favoured than the triangular lattice (Fig
~\ref{fig:lim:pot}(d)). The average density plot
(Fig.~\ref{fig:lim:lambda}(f)) shows sharp contours corresponding
to a rectangular structure.

Therefore, once again, we observe that as a function of the
location of the minimum of the attractive part of the
inter-particle potential, the system switches from one kind of
crystalline structure (triangular) to another ordered structure
(rectangular) via a modulated liquid phase.

\section{Conclusions}

In summary, in this paper we have investigated the effect of an
external laser field, periodically modulated in one dimension, on
a system of colloidal particles confined to two-dimensions and
interacting via a potential that includes a short-range attraction
component. The presence of this attractive interaction introduces
a new length scale (corresponding to the minimum of the attractive
pair potential) which can be incommensurate with the wavelength of
the potential due to the externally applied laser field. We find
that the competition between the two incommensurate length-scales
results in the phenomenon of {\it reentrant crystallization} at
high field strengths. In both the models we have studied, we find
that the crystalline phase attained by the particles at low field
strengths melts into a modulated liquid when the strength of the
field is increased. However, it eventually crystallizes again, at
higher field strengths, due to suppression of the particle motion
in the direction of modulation of the applied potential. This
phenomenon is opposite to the situation when the attractive
interaction is absent, where one observes only re-entrant melting.
We have also observed, at least in  one of the models, that
depending upon the position of the minimum of the attractive
potential, one can get different phases when both the particle
density and the laser potential are kept constant.

We hope that our simulations will motivate experiments to observe
the effect of the laser field in two-dimensional colloidal
mixtures. Also, in the context of colloidal aggregation, our
simulations show that structures with square symmetry in the
aggregated phase can be stabilized in the presence of substrates
formed by laser intensity fringes. We have also shown that other
novel structures can be formed by tuning various parameters of the
colloidal system. Further theoretical studies to understand the
phenomena observed in our simulations in greater detail would be
interesting to pursue.

\section{Acknowledgement}
PC would like to thank SERC (IISc) for providing the necessary computation facilities and 
JNCASR for providing financial support.

\newpage


\newpage

\newpage

\newpage

\newpage

\newpage

\newpage

\end{document}